\documentclass[prl,twocolumn,superscriptaddress]{revtex4}

\usepackage{tabularx}
\usepackage{float}
\usepackage{bm}
\usepackage{graphicx}
\usepackage{amsmath}
\usepackage{mathtools}
\usepackage{color}
\usepackage{mathrsfs}
\usepackage{hyperref}

\usepackage{subfiles}

\begin{document}

\title{Anomalous Exciton Hall effect}

\author{V. K. Kozin}
\affiliation{Science Institute, University of Iceland, Dunhagi-3, IS-107 Reykjavik, Iceland}
\affiliation{ITMO University, Kronverkskiy prospekt 49, Saint Petersburg 197101, Russia}

\author{V. A. Shabashov}
\affiliation{ITMO University, Kronverkskiy prospekt 49, Saint Petersburg 197101, Russia}
\affiliation{
St. Petersburg Academic University of the Russian Academy of Sciences, 194021 St. Petersburg, Russia}

\author{A. V. Kavokin}
\affiliation{Westlake University, 18 Shilongshan Road, Hangzhou 310024, Zhejiang Province, China}
\affiliation{Institute of Natural Sciences, Westlake Institute for Advanced Study, 18 Shilongshan Road, Hangzhou 310024, Zhejiang Province, China}
\affiliation{School of Physics and Astronomy, University of Southampton, Highfield, Southampton SO17 1BJ, United Kingdom}
\author{I. A. Shelykh}
\affiliation{Science Institute, University of Iceland, Dunhagi-3, IS-107 Reykjavik, Iceland}
\affiliation{ITMO University, Kronverkskiy prospekt 49, Saint Petersburg 197101, Russia}

\begin{abstract}
It is well known that electrically neutral excitons can still be affected by crossed electric and magnetic fields that make them move in a direction perpendicular to both fields. 
We show that a similar effect appears in the absence of external electric fields, in the case of scattering of an exciton flow by charged impurities in the presence of the external magnetic field. As a result, the exciton flow changes the direction of its propagation that may be described in terms of the Hall conductivity for excitons. We develop a theory of this effect, which we refer to as the anomalous exciton Hall effect, to distinguish it from the exciton Hall effect that arises due to the valley selective exciton transport in transition metal dichalcogenides. According to our estimations, the effect is relatively weak for optically active or bright excitons in conventional GaAs quantum wells, but it becomes significant for optically inactive or dark excitons, because of the difference of the lifetimes. This makes the proposed effect a convenient tool for spatial separation of dark and bright excitons.  
\end{abstract}

\maketitle

\textit{Introduction.} Thomas and Hopfield \cite{Thomas1961} were the first to point out that excitons propagating in the presence of an external magnetic field orthogonal to their velocity acquire stationary dipole polarisation perpendicular to both the magnetic field and their propagation direction. This is the manifestation of the magnetic Stark effect for excitons that was experimentally evidenced in a variety of semiconductor systems \cite{Yakovlev,Chen,Masha}. This effect is caused by the Lorentz force that pulls an electron and a hole apart if an exciton as a whole particle moves in the presence of a magnetic field. Imamoglu \cite{Imamoglu1996,Imamoglu2017} pointed out that once an exciton is placed in crossed electric and magnetic fields, it starts moving as a whole in the direction perpendicular to the directions of both fields, that leads to the renormalization of the excitonic dispersion in quantum wells or two-dimensional semiconductor crystals. The dynamics of electrically neutral quantum liquids in the presence of crossed electric and magnetic fields was studied by Shevchenko~\cite{Shevchenko1978,ShevchenkoPRL, Shevchenko2020}. Onga \textit{et al.}~\cite{Onga} have recently reported the experimental observation of an exciton Hall effect in atomically thin layers of MoS$_2$ that manifests itself in the appearance of an off-diagonal exciton conductivity in the presence of a magnetic field. The effect is caused by the strong spin-valley locking in monolayer transition metal dichalcogenides (TMDs). 

Here we predict an anomalous exciton Hall effect that is independent of the exciton spin structure. We show that in conventional GaAs quantum wells containing charged impurities, an exciton flow may be reoriented in the real space due to the combined effect of the local electrostatic potential created by charged impurities and the orbital effect of a magnetic field applied in normal to the plane direction. Conceptually, this effect is similar to the cross-field effect proposed by Imamoglu \cite{Imamoglu1996,Imamoglu2017} and it manifests itself in a very similar phenomenology to the exciton Hall effect studied by Onga \textit{et al.}~\cite{Onga}, however, it is different from both above mentioned effects as neither external electric field nor spin-valley locking are required in our case. To distinguish from the previous studies and emphasize the similarity with the anomalous Hall effect (AHE), we refer to the effect we study as an anomalous exciton Hall effect. 

We argue that the effect may have a significant magnitude in fluids of optically inactive, dark excitons due to their long lifetimes, and it is much weaker for short-living bright excitons in conventional GaAs-based quantum wells. This makes the anomalous exciton Hall effect a powerful tool for spatial separation of dark and bright excitons.

\textit{Synthetic gauge fields.} 
In condensed matter physics, gauge fields are ubiquitous. The best known example is a magnetic field $\mathbf{B}$, which can be introduced into a single particle Hamiltonian by substitution
$\hat{p}_{i}\rightarrow\hat{p}_{i}-q A_{i}$, with $q$ being the electric charge of the particle, $A_{i}$  being components of the vector potential, $\mathbf{B}=\nabla\times \mathbf{A}$. The presence of a magnetic field dramatically modifies the properties of the system, and leads to such fundamental phenomena as the quantum Hall and Aharonov-Bohm effects. For neutral particles with $q=0$, a magnetic field does not affect the orbital degree of freedom directly, and thus can not be considered as a real gauge field. However, if a particle possesses  internal degrees of freedom, such as spin, polarization, or internal set of energy levels, creation of so called synthetic gauge fields becomes feasible. In particular, for cold atoms both Abelian and non Abelian gauge potentials can be engineered by resonant drive of the system with spatially inhomogeneous laser beams \cite{Dalibard2011}. For photons, the methods to create synthetic gauge fields include dynamic modulation \cite{Fang2012}, use of coupled optical resonators \cite{Hafezi2011}, engineering lattices with strain \cite{Rechtsman2013}, or reciprocal metamaterials \cite{Liu2019}.

In condensed matter physics, the typical example of an electrically neutral quasiparticle is an exciton. Excitons govern the optical response of semiconductor materials at low and, in many cases, at high temperatures \cite{Wang2018}. The recent study of the electric field effect on the gauge fields for exciton strongly coupled to light (exciton-polaritons) \cite{Imamoglu2017} showed the high potentiality of the gauge field approach to the description of exciton and polariton dynamics in the presence of a magnetic field. In this Letter we demonstrate how the combination of the magnetic Stark effect \cite{Thomas1961} with excitonic scattering by an impurity potential leads to the anomalous exciton Hall effect link to the appearance of an effective U(1) gauge field acting on the motion of the exciton center-of-mass.

\textit{Phenomenological model.} 
We start with a simplified phenomenological model, assuming that the motion of the exciton center-of-mass is decoupled from the relative motion of the electron-hole pair. We consider a 2D exciton confined in the $xy$-plane and subject to the external magnetic field directed along the $z$-axis. 
If the exciton center of mass is characterised by a non-zero momentum $\mathbf{k}\neq0$, the magnetic field acting on the electron and hole would dipole-polarize the exciton in the direction perpendicular to $\mathbf{k}$, so that the electric dipole moment of the exciton reads
$\langle\hat{\mathbf{d}}\rangle=f(B)[\mathbf{e}_z\times\mathbf{k}]$, where  $f(B)$ is a function of the magnetic field, which depends linearly on $B$ at weak fields, but becomes inversely proportional to $B$ at the large fields in the magneto-exciton regime \cite{Lerner1978,Butov2001,Lozovik2002}. At small magnetic fields one can find $f(B)$ by passing to the center-of-mass reference frame, where the magnetic field is absent, but an electric field $\mathbf{E}^{\prime}=[\hbar \mathbf{k}\times\mathbf{B}]/M$ appears, here $M$ is the exciton mass. If this electric field is weak, then $\langle\hat{\mathbf{d}}\rangle=-e\langle\mathbf{r}\rangle=\alpha\mathbf{E}^{\prime}=\alpha[\hbar \mathbf{k}\times\mathbf{B}]/M$, where  $\alpha=21a_{\text{B}}^3 4\pi\varepsilon_0\varepsilon/128$ is the 2D exciton polarizability~\cite{Pedersen}, thus $f(B)=-\alpha\hbar B/M$. At strong magnetic fields, in the magneto-exciton regime, the dipole moment is given by~\cite{Lozovik2002} $\langle\hat{\mathbf{d}}\rangle=-e\langle\mathbf{r}\rangle=-e[\mathbf{e}_z\times\mathbf{k}]l_B^2$, where $l_{B}=\sqrt{\hbar/(eB)}$ is the magnetic length, thus $f(B)=-\hbar/B$.

\begin{figure}[h!]
	\includegraphics[width=1\linewidth]{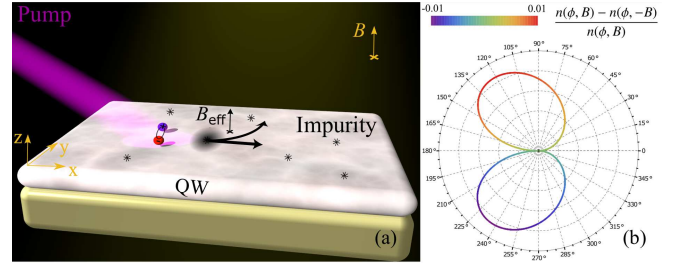}
	\caption{(a) The sketch of the system under consideration. Excitons created by the optical pump travel in the plane of a disordered quantum well in the presence of a uniform orthogonal magnetic field $\mathbf{B}$. The magnetic field induces an in-plane dipole moment of excitons which leads to the asymmetric scattering of excitons by impurities. This problem can be mapped to the scattering of a charged particle by an impurity in the presence of the position-dependent synthetic magnetic field $\mathbf{B}_\text{eff}(\mathbf{R})$, depicted by the grey domain around the scatterer in the figure. (b) Polar plot of $(n(\varphi,B)-n(\varphi,-B))/n(\varphi,B)$, where $n(\varphi,B)$ is the concentration of excitons, propagating in the direction given by the in-plane polar angle $\varphi$, plotted at $B=1$~T, see details in Supplementary Material~\cite{supp_mat}.
	}
    \label{fig:fig1_sketch}
\end{figure}

The presence of impurities and fluctuations of the doped quantum well width leads to the appearance of the scattering potential $U_{\text{sc}}(\mathbf{R})$ for excitons, here $\mathbf{R}$ denotes the position of the exciton center-of-mass. Moreover, it generates some non-zero in-plane distribution of the electric field $\mathbf{E}(\mathbf{R})$, which can affect the exciton dipole moment (see Fig.~\ref{fig:fig1_sketch} (a)).
Using these assumptions we write down the model Hamiltonian of the system as:
\begin{equation}
\hat{H}^{\text{dip}}=\frac{\hat{\mathbf{p}}^2}{2M}+\hat{V}^{\text{dip}}=\frac{\hat{\mathbf{p}}^2}{2M}+U_{sc}(\mathbf{R})-\frac{1}{2}(\hat{\mathbf{d}}\cdot\mathbf{E}+\mathbf{E}\cdot\hat{\mathbf{d}}),
\end{equation}
where $\hat{\mathbf{d}}=f(B)[\mathbf{e}_z\times\hat{\mathbf{k}}]$, $M$ is the exciton mass, and $\hat{\mathbf{p}}=\hbar\hat{\mathbf{k}}$ is the exciton center-of-mass momentum operator. Note, that as an exciton is a composite particle, there is no simple straightforward relation between the scattering potential and the electric field $\mathbf{E}=\mathbf{E}(\mathbf{R})$, produced by the scatterer.  After some algebra this can be cast in form $\hat{H}^{\text{dip}}=\left(\hat{\mathbf{p}}-e\mathbf{A}_{\text{eff}}(\mathbf{R})\right)^2/(2M)+U_{\text{eff}}(\mathbf{R})$, where $\mathbf{A}_{\text{eff}}(\mathbf{R})=M f(B)(\hbar e)^{-1}\left(\mathbf{e}_xE_y(\mathbf{R})-\mathbf{e}_y E_x(\mathbf{R})\right)$ is a synthetic vector potential, corresponding to the magnetic field $\mathbf{B}_{\text{eff}}(\mathbf{R})=-\mathbf{e}_z M f(B)(\hbar e)^{-1}\operatorname{div}\mathbf{E}(\mathbf{R})$. $U_{\text{eff}}(\mathbf{R})=U_{\text{sc}}(\mathbf{R})-M f^2(B)(2\hbar^2)^{-1}\mathbf{E}^2(\mathbf{R})$
is the effective scalar potential. One can see that for the appearance of a non-trivial synthetic gauge field, two conditions need to be fulfilled: (a) $B\neq 0$ and (b) the presence of an inhomogeneous electric field. 

We consider now the scattering of an exciton by a single radially symmetrical impurity with $U_{sc}(\mathbf{R})=U_{sc}(R)$ and $\mathbf{E}(\mathbf{R})=E(R)\mathbf{R}/R$. We obtain the following elastic scattering matrix elements between the exciton states with the center-of-mass momenta $\mathbf{k}$ and $\mathbf{k}^{\prime}$ ($|\mathbf{k}|=|\mathbf{k}^{\prime}|$)
\begin{equation}
V^{\text{dip}}_{\mathbf{k},\mathbf{k}^{\prime}} = U_{\text{sc}}(\Delta\mathbf{k})+  
\dfrac{i\hbar f[\mathbf{k}^{\prime} \times\mathbf{k}]_z}{ 4\pi|\Delta\mathbf{k}|} \int\limits_{0}^{\infty} \mathrm{d}R^2 E J_1 (|\Delta\mathbf{k}|  R).\label{ScatteringElementSimpModel}
\end{equation}
Here we performed the integration over the polar angle, which yielded the Bessel functions, $\Delta \mathbf{k}=\mathbf{k}^{\prime}-\mathbf{k}$ is the transferred momentum, the arguments of $f(B)$ and $E(R)$ are omitted for brevity.  The first  term $U_{\text{sc}}(\Delta\mathbf{k})\equiv(2\pi)^{-1}\int\limits_{0}^{\infty}R\mathrm{d}R U_{sc}(R)J_0(|\Delta\mathbf{k}| R)$ is real and it describes the normal symmetric scattering of an exciton by an impurity, while the second imaginary term accounts for the breaking of the time reversal symmetry by the magnetic field (as $V^{\text{dip}}_{\mathbf{k},\mathbf{k}^{\prime}}\neq V^{\text{dip}}_{-\mathbf{k}^{\prime},-\mathbf{k}}$).
This term gives rise to the asymmetric scattering (analogous to skew scattering) of excitons by the impurities and thus leads to the excitonic analogue of the anomalous Hall effect. Thus, the physical mechanism of the proposed effect is similar to the AHE~\cite{AHE_review}, where the role of spin-orbit coupling is replaced by momentum dependent exciton dipole polarization.

\textit{Microscopic theory}. 
Now we proceed with a full microscopic model of exciton-impurity scattering in the presence of a magnetic field that accounts for the coupling of electron-hole relative motion and the exciton center-of-mass motion. 

The Hamiltonian $\hat{H}_{\text{rel}}$ of the relative motion of an $e-h$ pair characterized by the center-of-mass momentum $\hbar \mathbf{k}$ in the presence of a perpendicular magnetic field $\mathbf{B}=(0,0,B)$ has the form~\cite{Lerner1978, GorkDzyal1968, Kyriienko2012}
\begin{equation}
\label{eq:hamilt_rel}
\begin{aligned}
\hat{H}_{\text{rel}}=&-\frac{\hbar^{2}}{2 \mu} \nabla_{\mathbf{r}}^{2}-\frac{i \hbar e}{2 }\left(\frac{1}{m_{e}}-\frac{1}{m_{h}}\right)\mathbf{B}\cdot\left[\mathbf{r} \times \nabla_{\mathbf{r}}\right]+ \\
&\frac{e^{2} B^{2}}{8 \mu} r^{2}+\frac{e \hbar}{M} \mathbf{B} \cdot[\mathbf{r} \times \mathbf{k}]-\frac{e^{2}}{4\pi\varepsilon_0\varepsilon|\mathbf{r}|}+\frac{\hbar^2\mathbf{k}^2}{2M},
\end{aligned}
\end{equation}
where $\mathbf{r}=\mathbf{r}_{e}-\mathbf{r}_{h}$ is the relative $e-h$ coordinate, and $\mu^{-1}$ $=m_{e}^{-1}+m_{h}^{-1} .$ Deriving this expression we have taken advantage of the existence of an exact integral of motion, namely the magnetic center-of-mass momentum~\cite{GorkDzyal1968}, defined by the operator
$\hbar \hat{\mathbf{k}}=-i \hbar \nabla_{\mathbf{R}}-e \mathbf{A}(\mathbf{r})$,
where $\mathbf{R}=\left(m_{e} \mathbf{r}_{e}+m_{h} \mathbf{r}_{h}\right) / M$ is the center-of-mass coordinate, $M=m_{e}+m_{h},$ and the vector potential is taken in the symmetrical gauge $\mathbf{A}(\mathbf{r})=\mathbf{B} \times \mathbf{r} / 2 .$ The exciton wave function has the form $\Psi_{\mathbf{k}}(\mathbf{R}, \mathbf{r})=\exp \left\{i \frac{\mathbf{R}}{\hbar}\left[\hbar\mathbf{k}+e \mathbf{A}(\mathbf{r})\right]\right\} \Phi_{\mathbf{k}}(\mathbf{r})$, where $\Phi_{\mathbf{k}}(\mathbf{r})$ is the corresponding eigenstate of the Hamiltonian above.
An important point is that the wave function $\Phi_{\mathbf{k}}$ of the relative motion depends on the center-of-mass momentum $\hbar\mathbf{k}$~\cite{GorkDzyal1968} i.e., the relative motion and the center-of-mass motion are coupled. The scattering matrix elements $
    V_{\mathbf{k}, \mathbf{k}^{\prime}}=\langle\Psi_{\mathbf{k}}|\hat{V}| \Psi_{\mathbf{k}^{\prime}}\rangle$ between the exciton states with the center-of-mass momenta $\mathbf{k}$ and $\mathbf{k}^{\prime}$ in the external potential $\hat{V}=V_{e}(\mathbf{r}_e)+V_{h}(\mathbf{r}_h)$ are given below.

\textit{Weak magnetic fields.}  In the weak-field
limit, $l_{B} \gg a_{\text{B}},$ the scattering matrix elements can be found analytically, here 
$a_{\text{B}}=4\pi\varepsilon_0\varepsilon\hbar^2/(\mu e^2)$ is the Bohr radius (the $1s$ exciton radius is $a_{\text{B}}/2$). We shall neglect excitonic transitions to the excited states of internal $e-h$ motion, i.e. the center-of-mass momentum $|\mathbf{k}|\ll a_{\text{B}}^{-1}$. The ground-state wave function $\Phi_{\mathbf{k}}(\mathbf{r})$ is calculated in a magnetic field using the perturbation theory.
The corresponding scattering matrix elements $V_{\mathbf{k}, \mathbf{k}^{\prime}}$ are obtained in Ref.~\cite{Arseev1998} and read as follows
\begin{align}
\label{eq:V_matrix_elements}
&V_{\mathbf{k},\mathbf{k}^{\prime}}=\tilde{V}_{e}(\Delta \mathbf{k})\mathcal{F}_{e}(\Delta\mathbf{k})+\tilde{V}_{h}(\Delta \mathbf{k})\mathcal{F}_{h}(\Delta\mathbf{k})+\nonumber\\
&i[\mathbf{k}^{\prime}\times\mathbf{k}]_z a_{\text{B}}^2\left(\frac{a_{\text{B}}}{l_B}\right)^2\left(\tilde{V}_e(\Delta\mathbf{k})\alpha_e-\tilde{V}_h(\Delta\mathbf{k})\alpha_h\right).
\end{align}
Here $\tilde{V}_{j}(\mathbf{k})$ is the two-dimensional Fourier transform of the potentials $V_{j}(\mathbf{r}) \quad(j=e, h)$,
$\mathcal{F}_{e(h)}(\Delta\mathbf{k})=\left[1 - 3 a_{\text{B}}^2 m_{h(e)}^2 (\Delta \mathbf{k})^2/(32M^2)+\beta_{e(h)}(\Delta \mathbf{k})^{2} a_{\text{B}}^{2}\left(a_{\text{B}}/\ell_{B}\right)^{4}\right]$, and
$\Delta \mathbf{k}=\mathbf{k}^{\prime}-\mathbf{k}$ is the transferred momentum. In the derivation above only the 
terms of up to the second order in $B$ and the lowest order in $|\mathbf{k}| a_{\text{B}}$ are taken into account. The expressions for the dimensionless constants $\alpha_{e(h)}$ and $\beta_{e(h)}$ for a 2D Wannier-Mott exciton can be found in~\cite{Arseev1998} (we corrected typos in the original formulas):
$\beta_{e(h)}=4^{-6} M^{-2}\left(105 m_{h(e)}^{2}-159 \mu^{2}/2\right)$
and
$\alpha_{e(h)}=-2 m_{h(e)}\kappa/M $, $\kappa=-21\mu/(16^2 M)$. Note that $\beta_{e}, \beta_{h}>0$ are positive, therefore, the exciton scattering cross-section increases with
$B$ when $l_{B} \gg a_{\text{B}}$. An important point is that the time-reversal symmetry is broken $V_{\mathbf{k}, \mathbf{k}_{1}} \neq V_{-\mathbf{k}_{1},-\mathbf{k}}$,
and the structure of $V_{\mathbf{k},\mathbf{k}^{\prime}}$  resembles its counterpart~(\ref{ScatteringElementSimpModel}) from the simplified phenomenological model. The ground-state exciton energy spectrum is given by
\begin{equation}
\label{eq:exciton_spectrum}
\epsilon(\mathbf{k})=\frac{\hbar^{2} \mathbf{k}^{2}}{2 M}\left[1-2|\kappa|\left(\frac{a_{\text{B}}}{l_{B}}\right)^{4}\right]-\epsilon_{0}\left[1-\left(\frac{l_{2}}{l_{B}}\right)^{4}\right],
\end{equation}
where the parameter $l_{2}=(3/128)^{1/4} a_{\text{B}}$ determines the diamagnetic shift. The first term stands for the center-of-mass kinetic energy, whereas the second term represents the binding energy.
\begin{figure}[h!]
	\center{\includegraphics[width=1\linewidth]{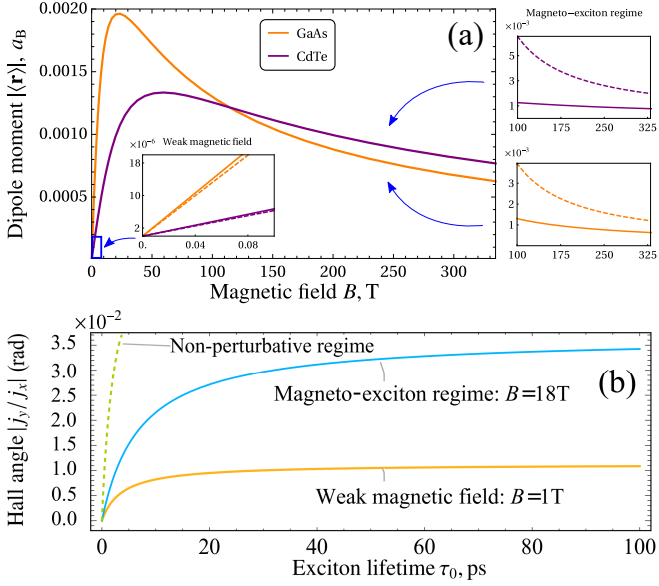}}
	\caption{
(a) The dependence of the dipole moment (absolute value $|\langle\mathbf{r}\rangle|$) on the magnetic field strength for GaAs and CdTe quantum wells. The solid lines are obtained from the numerically found eigenstates of Eq.~(\ref{eq:hamilt_rel}), whereas the dashed lines are obtained, using the perturbation theory.   The parameters of the plot are typical of GaAs quantum wells: the effective electron and hole masses $m_e = 0.067 m_{e0}$, $m_h = 0.5\, m_{e0}$, the dielectric constant of the quantum well $\varepsilon = 12.5$ 
The exciton momentum corresponds to the wavelength of the absorbed photon $\lambda=860$~nm as $|\mathbf{k}| =2\pi\sin(\pi/2-\pi/8)/\lambda$, where the angle of incidence is set to $\pi/2-\pi/8$. The linear part of the curve describes the weak magnetic field limit, whereas the decaying tail corresponds to the magneto-exciton regime.
(b) The Hall angle in a GaAs quantum well as a function of the exciton lifetime in two regimes: weak fields $a_{\text{B}}/l_B\ll 1$ and the magneto-exciton regime $a_{\text{B}}/l_B\gg 1$. The expected behavior in the non-perturbative regime $a_{\text{B}}/l_B\sim 1$, where the Hall angle is expected to be the largest, is shown with the dashed curve.}
\label{fig:dipole_moment}
\end{figure}

\textit{Magneto-exciton regime.} In the opposite regime, where the magnetic field is strong (i.e. $a_{\text{B}}\gg l_B$), one can neglect the Coulomb interaction term in the Hamiltonian~(\ref{eq:hamilt_rel}) and the exciton dynamics is governed by the magnetic field solely. In this regime, which is often referred to as the magneto-exciton regime, one can calculate the dispersion of the ground state of the system, treating the $e-h$ Coulomb potential as a perturbation~\cite{magneto_excitons_dzyubenko, Bychkov1980}
\begin{equation}
\label{eq:magneto-exciton_dispersion}
    \epsilon(\mathbf{k})=\frac{1}{2}\hbar\omega_{\text{C}}-\sqrt{\frac{\pi}{2}}\frac{e^2}{4\pi\varepsilon_0\varepsilon l_B} e^{ -\frac{k^{2} l_{B}^{2}}{4}} I_{0}\left(\frac{k^{2} l_{B}^{2}}{4}\right),
\end{equation}
where $I_0$ is the modified Bessel function. The corresponding impurity scattering matrix elements are given by 
\begin{align}
\label{eq:V_matrix_elements_strong}
V_{\mathbf{k},\mathbf{k}^{\prime}}=& \tilde{V}_{e}(\Delta \mathbf{k}) \exp \left(-\frac{i}{2}\left[\mathbf{k}^{\prime} \times \mathbf{k}\right]_{z} l_{B}^{2}-\frac{\Delta \mathbf{k}^{2} l_{B}^{2}}{4}\right)+\nonumber\\
&\tilde{V}_{h}(\Delta \mathbf{k}) \exp \left(+\frac{i}{2}\left[\mathbf{k}^{\prime} \times \mathbf{k}\right]_{z} l_{B}^{2}-\frac{\Delta \mathbf{k}^{2} l_{B}^{2}}{4}\right),
\end{align}
here $\omega_{\text{C}}=eB/\mu$ is the cyclotron frequency.

\textit{Scattering rates}. The 
scattering $T$-matrix can be defined as $T_{\mathbf{k},\mathbf{k}^{\prime}}=\langle\Psi_\mathbf{k}\lvert\hat{V}\rvert\tilde{\Psi}_{\mathbf{k}^{\prime}}\rangle$, where $\hat{V}=V_{e}\left(\mathbf{r}_{e}\right)+V_{h}\left(\mathbf{r}_{h}\right)$ is the impurity potential operator, $\lvert\Psi_\mathbf{k}\rangle$ is the eigenstate of the free Hamiltonian $\hat{H}_0$, describing a 2D exciton in a magnetic field, and $\lvert \tilde{\Psi}_{\mathbf{k}^{\prime}}\rangle$ is
the eigenstate of the full Hamiltonian $\hat{H}=\hat{H}_0+\hat{V}$. The $T$-matrix 
satisfies the Lippmann-Schwinger equation 
\begin{equation}
\label{eq:Lippmann_Schwinger}
    T_{\mathbf{k}, \mathbf{k}^{\prime}}=V_{\mathbf{k}, \mathbf{k}^{\prime}}+\int \frac{d^2\mathbf{g}}{(2\pi)^2} \frac{V_{\mathbf{k}, \mathbf{g}} T_{\mathbf{g}, \mathbf{k}^{\prime}}}{E-\epsilon(\mathbf{g})+i 0},
\end{equation}
where $\epsilon(\mathbf{g})$ is the dispersion of the bare Hamiltonian $\hat{H}_0$ and it is given by Eq.~(\ref{eq:exciton_spectrum}) and Eq.~(\ref{eq:magneto-exciton_dispersion}) for the two regimes under consideration, respectively, and $E$ is the energy eigenvalue corresponding to $\lvert\tilde{\Psi}_{\mathbf{k}^{\prime}}\rangle$. Two contributions can be distinguished in the square modulus of the $T$-matrix: $\nu_0^{2}\left|T_{\mathbf{k}, \mathbf{k}^{\prime}}\right|^{2}=\mathcal{G}_{\mathbf{k},\mathbf{k}^{\prime}}+\mathcal{J}_{\mathbf{k},\mathbf{k}^{\prime}}$,
here $\mathcal{G}_{\mathbf{k},\mathbf{k}^{\prime}}\equiv\mathcal{G}(\theta)=\mathcal{G}(-\theta)$, $\mathcal{J}_{\mathbf{k},\mathbf{k}^{\prime}}\equiv\mathcal{J}(\theta)=-\mathcal{J}(-\theta)$ are dimensionless symmetric and asymmetric contributions to the scattering rate, respectively, $\theta=\varphi-\varphi^{\prime}$ is the scattering angle, $\varphi, \varphi^{\prime}$ are the polar angles of $\boldsymbol{k}, \boldsymbol{k}^{\prime},$ and $\nu_0=M/(2 \pi \hbar^{2})$
is the 2D density of states of free particles with parabolic dispersion. 
The density of states is defined as
$\nu(\mathbf{k})=\left|\partial\epsilon(\mathbf{k})/\partial k\right|^{-1} k/(2\pi)$. Here we restrict ourselves to the case of elastic scattering $|\mathbf{k}|=|\mathbf{k}^{\prime}|$.
It is the asymmetric part $\mathcal{J}_{\mathbf{k},\mathbf{k}^{\prime}}$ of exciton scattering by
impurities that gives rise to the Hall current. The properties of $\mathcal{J}_{\mathbf{k},\mathbf{k}^{\prime}}$ are discussed in details in Ref.~\cite{Denisov2017,DenisovPRL206,Denisov2018}.

The elastic scattering rate  $W_{\mathbf{k},\mathbf{k}^{\prime}}$ from $\mathbf{k}^{\prime}$ to $\mathbf{k}$ state is expressed with use of the Fermi golden rule
$W_{\mathbf{k}, \mathbf{k}^{\prime}}=2 \pi\hbar^{-1} n_{\text{imp}}\left|T_{\mathbf{k}, \mathbf{k}^{\prime}}\right|^{2} \delta\left(\epsilon_{\mathbf{k}}-\epsilon_{\mathbf{k}^{\prime}}\right)$,
where $n_{\text{imp}}$ is the surface density of impurities. The presence of the magnetic field breaks the time inversion symmetry in the problem and leads to the disbalance of scattering rates $W_{\mathbf{k},\mathbf{k}^{\prime}}$ and $W_{\mathbf{k}^{\prime},\mathbf{k}}$, which is why a non-zero Hall contribution $\mathcal{J}_{\mathbf{k},\mathbf{k}^{\prime}}$ emerges.

Let us assume that the impurity potential is described by the Coulomb potential~\cite{FraizzoliImpurity1990} $V_e(r)=-V_h(r) = -e q_{\text{imp}}/(4 \pi \varepsilon_0 \varepsilon r)$.
The Lippmann-Schwinger equation can not be treated perturbatively for such a potential in our system, which is why we solve this integral equation~(\ref{eq:Lippmann_Schwinger}) numerically. 

The developed formalism allows us to predict the anisotropy of an angular distribution of the exciton emission that would appear if a flow of excitons propagates in a plane of a doped quantum well in the presence of a magnetic field normal to the plane of the well (as Fig.~\ref{fig:fig1_sketch}~(a) shows schematically). The angular distribution of the exciton emission may be found as $n(\phi) = \int\limits_{0}^{\infty} n_{\mathbf{k}} \; \dfrac{kdk}{(2\pi)^2}$,
where $n_{\mathbf{k}}$ is the occupation numbers of exciton states  having wave vectors $\mathbf{k}$.
The normalised variation of this quantity due to the inversion of the orientation of the applied field is shown in Fig.~\ref{fig:fig1_sketch}~(b). For a detailed description of the relevant formalism we refer the reader to the Supplemental material ~\cite{supp_mat}. The observation of a predicted variation of the angular distribution of the excitonic emission can be considered as a smoking gun for the anomalous exciton Hall effect. We note, that dark excitons contribute very little to the intensity of photoluminescence, while they strongly contribute to its blue shift~\cite{CombescotEPL2014,RapaportPNAS2019}. Experimental measurement of the angular resolved blue-shift of the bright exciton photoluminescence peak might certify the presence of a flow of dark excitons.

Next, to obtain the Hall angle, we study the classical transport regime $|\mathbf{k} |l\gg 1$, where $l$ is the exciton mean
free path. We use the semiclassical Boltzmann equation 
\begin{equation}
\frac{d n_{\mathbf{k}}}{d t}=P_{\mathbf{k}}-\Gamma n_{\mathbf{k}}+\int\frac{d^2\mathbf{k}^{\prime}}{(2\pi)^2}(W_{\mathbf{k},\mathbf{k}^{\prime}} n_{\mathbf{k}^{\prime}}-W_{\mathbf{k}^{\prime},\mathbf{k}}n_{\mathbf{k}}),
\end{equation}
here $P_{\mathbf{k}}$ is the coherent pump term, $\Gamma=1/\tau_0$ is the particle decay rate. In the stationary regime,
performing the integration over the absolute value of $\mathbf{k}^{\prime}$, the Boltzmann kinetic equation can be rewritten as $P_{\mathbf{k}}=\Gamma n_{\mathbf{k}}-\int d \varphi^{\prime}\left(w_{\mathbf{k },\mathbf{k}^{\prime}} n_{\mathbf{k}^{\prime}}-w_{\mathbf{k}^{\prime}, \mathbf{k}} n_\mathbf{k}\right)$,
where $w_{\mathbf{k},\mathbf{ k}^{\prime}}=n_{\text{imp}}\nu_0/(\hbar\nu(\mathbf{k})^2)\left(\mathcal{G}_{\mathbf{k},\mathbf{k}^{\prime}}+\mathcal{J}_{\mathbf{k},\mathbf{k}^{\prime}}\right)
$ and the exciton density of states $\nu(\mathbf{k})=\left|\partial\epsilon(\mathbf{k})/\partial k\right|^{-1} k/(2\pi)$. Let us assume the in-plane wave vector $\mathbf{k}_0$ of the pump is pointing along the $x$-axis, i.e. $\mathbf{k}_0=(k_0,0)^T$, which implies $P_{\mathbf{k}}=P_0\delta(\mathbf{k}-\mathbf{k}_0)=(P_0/k)\delta(k-k_0)\delta(\varphi)$ and we note that this function is even with respect to $\varphi$. 
Assuming that only the dipole type of anisotropy of the momentum-space distribution function is significant, we represent $n_{\mathbf{k}}=n_{0}(k)+\delta n(\mathbf{k})$, where
$\delta n(\mathbf{k})=n_{+}(k) \cos \varphi+n_{-}(k) \sin \varphi$,    
$n_{0}(k)$ is the isotropic part of the distribution function which depends only on energy. 
Substituting this decomposition into the kinetic equation and integrating over $\varphi^{\prime}$ in the collision term, we obtain
\begin{align}
&0=\frac{P_0}{k}\delta(k-k_0)\delta(\varphi)+\cos \varphi\left(\Omega(k) n_{-}(k)-\frac{n_{+}(k)}{\tau(k)}\right)\nonumber\\
&-\sin \varphi\left(\Omega(k) n_{+}(k)+\frac{n_{-}(k)}{\tau(k)}\right)-\frac{n_\mathbf{k}}{\tau_0},
\end{align}
where $\tau(k)$ is given by the symmetric scattering term $\tau(k)^{-1}=n_{\text{imp}}\nu_0/(\hbar\nu(\mathbf{k})^2) \int_0^{2\pi} \mathcal{G}_{\mathbf{k},\mathbf{k}^{\prime}}(1-\cos{\theta}) d \theta$, here $\theta=\varphi-\varphi^{\prime}$ is the scattering angle.  The factor $\Omega(k)$ is governed by the asymmetric scattering term
$\Omega(k)=-n_{\text{imp}}\nu_0/(\hbar\nu(\mathbf{k})^2) \int_0^{2\pi} \mathcal{J}_{\mathbf{k},\mathbf{k}^{\prime}} \sin{\theta} d \theta$,
and it mixes the even $n_{+}(k)$ and odd $n_{-}(k)$ contributions to the density distribution, yielding a Hall current in the transverse $y$-direction. Using the orthogonality of $\sin{\varphi}$ and $\cos{\varphi}$, the kinetic equation is readily solved, yielding  $n_{-}(k)=-\Omega(k)\tau_{\text{tot}}(k) n_{+}(k)$. Here we introduced the total relaxation time $\tau_{\text{tot}}(k)=(\tau_0^{-1}+\tau(k)^{-1})^{-1}$. The Hall angle~\cite{SchwabHallAngle}, is defined as the ratio of the Hall current $j_y$ and the longitudinal current $j_x$, is $j_y/j_x=-\Omega(k_0)\tau_{\text{tot}}(k_0)$, where
$j_{x,y}=\int \hbar k_{x,y} \delta n(\mathbf{k})d^2\mathbf{k}/(2\pi)^2$. Note that Onga et al.~\cite{Onga} use a different definition. We estimate the Hall angle for the typical parameters of doped GaAs quantum wells $n_{\text{imp}}\approx10^{11}$ cm$^{-2}$, $k_0\approx7\cdot10^6$~m$^{-1}$ (see the caption of Fig.~\ref{fig:dipole_moment}), $\varepsilon=12.5$, $q_{\text{imp}} = e$,
and $\tau_0\approx 10$ ps (for bright excitons). At a weak magnetic field $B=1$~T the numerical solution of the Lippmann-Schwinger equation with the kernel~(\ref{eq:V_matrix_elements}), yields $\tau(k_0)\approx 3.8$ ps and $j_y/j_x=-\Omega(k_0)\tau_{\text{tot}}(k_0)\approx 0.8\%$. At increased concentration and a strong magnetic field $B=18$~T we solve the Lippmann-Schwinger integral equation numerically with the magneto-exciton kernel~(\ref{eq:V_matrix_elements_strong}), which yields $j_y/j_x=-\Omega(k_0)\tau_{\text{tot}}(k_0)\approx 1.8\%$.  One may expect, by looking at Fig.~\ref{fig:dipole_moment}(a), that at intermediate magnetic field strengths (about $B\approx10$~T) the Hall angle would be significantly larger, as the exciton dipole moment reaches its largest value in this non-perturbative regime, while still not switching to the magneto-exciton regime. We also note, that the range of magnetic fields corresponding to the largest dipole moment is easily achievable experimentally, which makes the observation of the predicted phenomenon realistic. The dependence of the Hall angle on the exciton lifetime is shown in Fig.~\ref{fig:dipole_moment}(b). Clearly, for dark excitons whose lifetime is significantly larger than the lifetime of bright excitons, the Hall angle is notably larger. This shows that the anomalous exciton Hall effect may be used as a tool for spatial separation of dark and bright excitons. 

\textit{Conclusions.} In conclusion, we demonstrated that the magnetic Stark effect for 2D excitons may lead to the emergence of an effective U(1) gauge field. This field can result in the excitonic analogue of the anomalous Hall effects. For the later we presented a detailed microscopic description of the scattering mechanism and analyzed the transport properties, showing that the effect can be observed experimentally in conventional GaAs quantum wells and that it is much stronger for dark than bright excitons.

\textit{Acknowledgments.} The work of V.K.K. related to numerical analysis of the magnitude of the predicted effects ws supported by Russian Science Foundation, Grant No. 19-72-
20120. I.A.S. thanks Ministry of Education and Science of Russian Federation, Project
14.Y26.31.0015, and Icelandic Research Fund (Rannis), project "Hybrid Polaritonics". AK acknowledges Project No. 041020100118 and Program 2018R01002 supported
	by Leading Innovative and Entrepreneur Team Introduction Program of the Zhejiang Province.

\end{document}